# AI Regulation and Capitalist Growth: Balancing Innovation, Ethics, and Global Governance

Vikram Kulothungan*, Priya Ranjani Mohan†, Deepti Gupta‡
*Capitol Technology University, 11301 Springfield Rd, Laurel, Maryland 20708, USA
†Independent Researcher, Rutgers University, New Brunswick, NJ 08901, USA
‡Dept. of Computer Information Systems, Texas A&M University - Central Texas, 1001 Leadership Pl, Texas 76549, USA
*vikramk1986@gmail.com, †priyamohan0630@gmail.com, ‡d.gupta@tamuct.edu

*Abstract*—Artificial Intelligence (AI) is increasingly central to economic growth, promising new efficiencies and markets. This economic significance has sparked debate over AI regulation: do rules and oversight bolster long term growth by building trust and safeguarding the public, or do they constrain innovation and free enterprise? This paper examines the balance between AI regulation and capitalist ideals, focusing on how different approaches to AI data privacy can impact innovation in AI-driven applications. The central question is whether AI regulation enhances or inhibits growth in a capitalist economy. Our analysis synthesizes historical precedents, the current U.S. regulatory landscape, economic projections, legal challenges, and case studies of recent AI policies. We discuss that carefully calibrated AI data privacy regulations—balancing innovation incentives with the public interest—can foster sustainable growth by building trust and ensuring responsible data use, while excessive regulation may risk stifling innovation and entrenching incumbents.

## I. INTRODUCTION

Artificial Intelligence (AI) is advancing rapidly, raising questions about how to harness its benefits for economic growth while managing its risks in a market-driven society. A central challenge is balancing innovation incentives with societal safeguards in a capitalist economy. Overly lenient approaches may overlook negative externalities such as bias and safety hazards, while excessively strict rules might hinder technological progress. This challenge introduces a contrasting views between proactive regulatory intervention and the laissez-faire principles traditionally associated with capitalist development. In the United States (U.S.), AI policy is at a critical juncture as regulators work to mitigate societal risks while fostering innovation and economic growth cornerstones of a capitalist system. Historical precedents show that early minimal oversight can stimulate growth but may later require regulation to address market failures and social harms. For instance, the Industrial Revolution's unfettered capitalism delivered unprecedented productivity but also led to monopolies and labor exploitation—prompting antitrust laws and labor protections [1] [2]. Similarly, the early Internet's growth was aided by light touch policies like Section 230 of the 1996 Communications Decency Act, which immunized online platforms from liability, and paved the way for tremendous innovation, the rise of social media giants and also challenges such as market concentration and content moderation [3].

This paper examines whether regulatory intervention in AI can serve as a catalyst for sustainable capitalist growth by establishing trust, promoting fairness, and mitigating risks, or if it might inadvertently slow down the entrepreneurial dynamism that fuels innovation. Drawing on insights from innovation economics and regulatory theory, we propose that well crafted, adaptable guardrails, emphasizing outcomes based standards and demonstrable harms can enhance long term economic growth and public trust. Conversely, rigid or overly prescriptive regulations may increase compliance costs and limit leadership in AI development. Our analysis, which integrates historical context, current U.S. policy review, economic and legal analysis, relevant case studies, and future scenario exploration, aims to identify optimal strategies for managing the interplay between AI regulation and capitalist growth while promoting both innovation and societal well-being. We strive to offer actionable insights that policymakers and industry leaders can leverage to foster responsible AI development. Ultimately, our goal is to balance progress with public trust, ensuring a more equitable and sustainable future.

The main contributions of this paper are as follows.

- We present an analysis of industry self-regulation and ethical AI governance, highlighting their economic implications.
- We examine constitutional and legal challenges, discussing their impact on AI regulation and innovation.
- We provide historical insights from technology regulations, supported by case studies to illustrate key lessons.
- We propose a principle-based AI regulatory framework that promotes ethical and innovative growth, designed to be adaptable, actionable, and scalable across various domains.

The remainder of this paper is organized as follows. Section II presents about industry self-regulation and ethical AI governance. We discuss economic implications due to AI in Section III. The constitutional and legal challenges are discussed in the Section IV. The Section V presents the historical lessons from technology regulations. The proposed framework is presented in Section VI. The conclusion and future work are discussed in Section VII.

## II. INDUSTRY SELF-REGULATION AND ETHICAL AI GOVERNANCE

Many stakeholders advocate for industry self-regulation as a flexible and immediate mechanism to guide AI development responsibly. Industry self-regulation refers to the voluntary adoption of guidelines, principles, and oversight mechanisms by companies without direct legal mandates. In recent years, leading AI firms have publicly committed to "Responsible AI" principles encompassing fairness, transparency, safety, privacy, and accountability to ensure that their technologies align with ethical norms. Since 2016, dozens of organizations, including at least 24 major AI companies, have pledged adherence to such principles either through public statements or by joining multi-stakeholder initiatives like the Partnership on AI [4]. The Partnership on AI [5], founded by companies such as Amazon, Apple, Facebook, Google, IBM, and Microsoft, is a notable consortium focused on developing collaborative benchmarks and best practices for AI.

Importantly, many companies have progressed beyond mere declarations to tangible implementation. Research indicates that several large technology firms have taken concrete steps to operationalize their AI ethics commitments [4]. For example, Google [6], Microsoft [7], and IBM [8] have each established internal AI ethics review processes and boards, and have invested in tools to detect bias and enhance algorithmic transparency [4]. These companies have also implemented extensive employee training on ethical AI and published resources such as AI Guidebook and open-source fairness toolkits to integrate ethical practices into their development lifecycle. Such measures suggest that, for the most committed firms, ethical AI governance is translating into substantive corporate practices rather than superficial "ethics washing." Moreover, many industry leaders recognize that self-regulation has limitations; several companies have publicly acknowledged the need for targeted government oversight in high-risk areas, advocating a balanced approach that combines voluntary efforts with formal regulation.

As AI governance has evolved, key milestones have shaped its trajectory from industry-led self-regulation to global policy frameworks and legal mandates. The timeline outlines major developments, starting with early ethical initiatives and corporate AI ethics programs [9] [10], followed by international agreements like the OECD AI Principles [11] and the UNESCO AI Ethics Recommendation [12]. The transition to hard law began with the drafting of the EU AI Act, and efforts continue toward global harmonization with initiatives like the G7 Hiroshima AI Process [13]. These milestones highlight the shift toward structured, accountable, and internationally coordinated AI governance to ensure ethical, transparent, and safe AI deployment worldwide. Proponents argue that industry is often best positioned to respond rapidly to technological changes and establish context-appropriate norms. Self-regulation can help fill gaps and set standards in the fast-moving AI sector, at least in the interim [14]. By voluntarily converging on ethical norms, companies may also help shape future formal regulations toward pragmatic, innovation-friendly approaches. Nevertheless, it is widely acknowledged that self-regulation alone may not fully mitigate risks or ensure accountability across all actors, particularly in a competitive market. Therefore, a broader governance ecosystem including neutral international frameworks is essential to complement industry efforts.

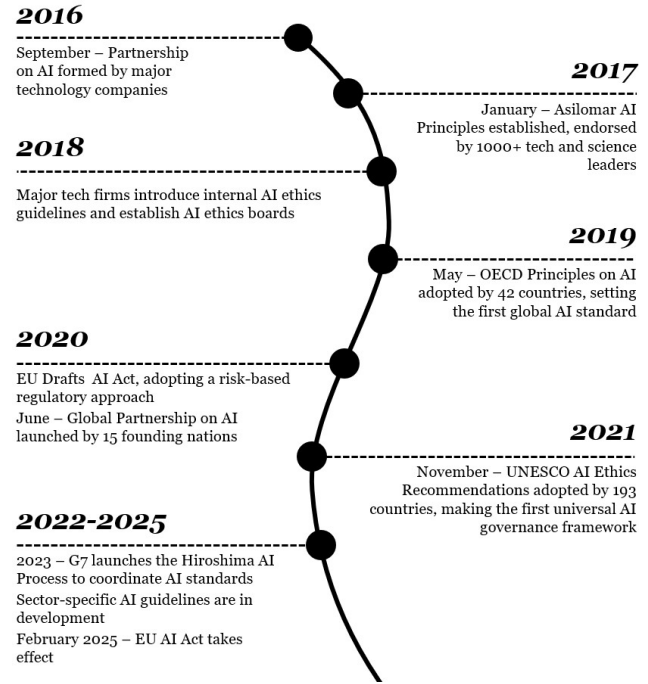

Fig. 1: AI Governance Timeline

## III. ECONOMIC IMPLICATIONS

Current economic evidence specifically relating to AI suggests that AI innovation is a significant driver of firm-level and macroeconomic growth. Projections for AI's macroeconomic benefits are sizeable. A widely cited analysis by Goldman Sachs (2023) estimated that generative AI could boost global GDP by about 7 percent (roughly 7 trillion dollars) over the next decade, largely by raising labor productivity [15]. Similarly, McKinsey Global Institute projected AI could add $13 to $26 trillion to annual global output by 2030. Such numbers have fueled optimism that AI will drive a new wave of capitalist growth, comparable to past general-purpose technologies. However, these gains are not automatic, and some economists urge caution. A study by MIT's Institute for Data, Systems, and Society offers a modest outlook: using task-level data, Acemoglu finds AI's likely contribution to U.S. productivity over the next 10 years might be only on the order of 0.5–1 percent of GDP [16]. In this research, many tasks are not easily automated by AI, and even where AI is adopted, the benefits may accrue narrowly unless complemented by broader changes. This "nontrivial but modest" macro impact suggests that policy choices will strongly influence whether AI's potential translates into broad economic growth or just concentrated gains. Indeed, Acemoglu and others warn that without intervention, AI could follow the path of past automation where productivity gains did not broadly raise wages due to weakened labor bargaining power [1].

A key question in the governance of AI is how regulation impacts innovation, especially in capitalist economies that prize entrepreneurship and growth. Economic theory and modeling offer insights into how different regulatory approaches can either encourage or dampen technological innovation. The relationship is complex: well-crafted rules can correct market failures and build trust in AI, potentially spurring adoption, whereas poorly designed or excessive regulations may impose costs that deter research and development [17]. The relationship between AI regulation and innovation is multifaceted, as summarized in Table II. In summary, both economic modeling and empirical evidence suggest that AI regulation and capitalist innovation need not be in opposition. Thoughtfully crafted governance can align private incentives with the public good, thereby creating a stable environment in which innovation thrives. The key is clarity, proportionality, and adaptability: regulations should clearly define unacceptable practices (thus reducing uncertainty for innovators), impose burdens commensurate with the level of risk, and remain flexible to update as technology and market conditions evolve.

## IV. Constitutional and Legal Challenges

AI regulation in the U.S. must also navigate constitutional liberties and federalism, adding another layer of complexity to the capitalist growth equation. Two constitutional areas stand out: First Amendment (free speech) and Fourth Amendment (search and seizure/privacy) concerns.

### A. First Amendment – AI-Generated Content and Speech

AI technologies are raising novel questions about speech rights. For example, generative AI can produce text, images, or deepfake videos that may be subject to content regulations. Laws that try to restrict or label AI-generated content (say, requiring watermarks on AI images or banning certain deepfake uses) must contend with free speech protections. Legal scholars suggest that AI-generated speech is generally protected by the First Amendment, not because the AI has rights, but because of the rights of the humans who develop and use the AI [19]. In other words, if an AI model produces some output, government restrictions on that output could infringe on the human creator's or user's right to speak or to receive information. Courts have long held that code can be speech; by extension, the outputs of algorithms may enjoy broad protection. That said, First Amendment protection is not absolute. The same exceptions that apply to human speech – defamation, fraud, obscenity, incitement, etc., would also apply to AI content [19]. For instance, an AI deepfake video that defames someone could be subject to libel laws. But if a regulation broadly banned "AI-generated political ads," it might be struck down as a content-based restriction on speech if it isn't narrowly tailored. Policymakers thus have to thread the needle: tackling harms like misinformation and fraud without violating free expression. The constitutional balance will influence how aggressively regulators can act against AI-driven disinformation, which in turn affects how freely companies can deploy AI content generation (an economic activity). A permissive legal environment for AI speech might maximize innovation (more tools, fewer restrictions), whereas stricter rules for labeling or limiting AI content might slow certain applications but protect public trust and democratic processes – arguably a prerequisite for stable growth.

### B. Fourth Amendment – Surveillance, Predictive Policing, and Privacy

AI can be used in surveillance and law enforcement, from facial recognition cameras to predictive policing algorithms. These technological capabilities can work against privacy and liberty rights. The Fourth Amendment requires that searches and seizures by the government be reasonable, often meaning law enforcement needs individualized suspicion (probable cause or reasonable suspicion) before intruding on privacy. Predictive policing – algorithms that analyze data to forecast crime hotspots or potential offenders – challenges this paradigm. If police deploy based on an algorithm's "hunch" rather than particularized evidence, it may erode the Fourth Amendment standard. Civil liberties organizations caution that "the Fourth Amendment forbids police from stopping someone without reasonable suspicion. Computer-driven hunches are no exception to this rule" [20]. In other words, an AI prediction alone shouldn't justify a stop-and-frisk; otherwise, we risk undermining constitutional protections with black-box algorithms. Courts have yet to fully grapple with these issues, but one federal appellate judge noted concerns that predictive tools could distort the reasonable suspicion analysis [21]. Similarly, AI-powered mass surveillance (like face recognition identifying people in public) could be seen as searches that currently evade warrant requirements. Regulations at state/local levels (e.g. some cities banning police use of face recognition) have emerged due to these privacy concerns. From an economic perspective, if privacy laws become too stringent (for example, severely limiting data collection), it could inhibit AI development in sectors like health and security. But if they are too lax, public backlash or legal challenges could also disrupt deployment. A balanced approach that clarifies permissible uses (such as requiring warrants for certain AI surveillance, or banning only the most intrusive practices) can provide guardrails without banning AI outright. Ensuring AI tools align with constitutional values is also critical for public trust: capitalist growth thrives under rule of law, and clear legal boundaries can actually foster innovation by defining what is off-limits and what is fair game.

In sum, constitutional and legal challenges don't make AI regulation impossible, but they require careful design. Regulations must respect free speech by targeting real harms (fraud, defamation) rather than suppressing speech broadly, and they must preserve privacy and due process, likely by en-hancing transparency and oversight of AI used by government. Getting this right enhances the legitimacy of AI technologies and prevents lengthy court battles that could delay beneficial deployments. It's part of ensuring that the capitalist innovation engine runs within the bounds of democratic values and the rule of law.

| Aspect | Key Insights | Implications |
|---|---|---|
| Economic Theory and Market Failures | Laissez-faire markets may underinvest in socially beneficial AI research. Firms may overinvest in profitable but socially harmful AI applications [18] | Regulations should correct market failures by incentivizing positive spillovers and limiting negative externalities |
| Empirical Evidence on AI Innovation | AI-driven firms experience higher revenue growth, employment, and productivity. AI adoption contributes to wage inequality, benefiting high-skilled workers [17] | Regulators must balance growth incentives with inclusivity measures to prevent market concentration and inequality |
| Balancing Regulation for Sustainable Growth | Antitrust policies and competitive AI markets encourage innovation by smaller players. Workforce retraining programs help distribute AI-driven productivity gains. A risk-based approach to AI regulation (e.g., stricter rules for high-risk applications, flexible policies for low-risk AI) [4] | Ensures long-term trust and stability in AI markets. Prevents monopolistic control and labor displacement backlash |
| AI Regulation as an Iterative Process | Companies adapt to regulations creatively, influencing the regulatory landscape. Continuous dialogue between policymakers and industry is essential | Regulation should be flexible, allowing adjustments based on real-world AI advancements and market responses |

TABLE I: Key Insights and Implications of AI Regulation

## V. HISTORICAL LESSONS FROM TECHNOLOGY REGULATION

Contemporary debates on AI governance benefit from historical analogies. Over the past century, several technological revolutions—such as those in aviation, biotechnology, and the internet—have raised comparable issues regarding innovation and regulation. An examination of how these technologies were governed provides evidence-based lessons for managing AI. Historical evidence suggests that balanced, adaptive regulatory approaches have often enabled emerging industries to thrive, whereas overly extreme or hasty interventions can have adverse effects. Key insights from these precedents include the value of industry initiative, the effectiveness of risk-based and incremental regulatory measures, and the importance of global cooperation.

In both biotechnology and internet domains, early regulatory approaches were targeted rather than blanket, thereby enabling these emerging technologies to flourish. For instance, during the early days of recombinant DNA technology in the 1970s, public concerns regarding potential biohazards led leading scientists to impose a voluntary moratorium on certain experiments and convene the Asilomar Conference in 1975 to develop safety guidelines. This act of scientific self-regulation, which produced the first "rulebook" for genetic engineering in collaboration with government observers and ethicists, resulted in a set of containment procedures that laboratories worldwide later adopted. By the time formal governmental regulations—such as NIH guidelines and subsequent genetic engineering laws—were introduced, the industry had already integrated many best practices, thereby easing compliance [22]. The Internet provides another instructive example. In the 1990s, as the Internet evolved from a research network into a global commercial platform, U.S. policymakers chose a "light-touch" regulatory strategy that allowed the private sector to take the lead [22]. By encouraging industry self-regulation and avoiding burdensome rules or taxes, the Internet's early governing structures—such as ICANN for domain names, the W3C for web standards, and the Internet Engineering Task Force for technical protocols were developed through multistakeholder consensus. This decentralized, open approach is widely regarded as a major factor in the Internet's rapid growth. Although later challenges, including online fraud, cybersecurity issues, and digital monopolies, necessitated more targeted regulatory interventions, the early emphasis on "permissionless innovation" drove significant investment and creativity. International cooperation is also essential for technologies that transcend national boundaries. Historical examples include the Chicago Convention of 1944, which established unified rules for international air travel [23]. These coordinated standards prevented a fragmented regulatory landscape and facilitated market expansion. In the context of AI, efforts such as UNESCO's Recommendation [24] and emerging ISO standards [25] underscore the importance of global coordination. Similar to telecommunications standards developed through bodies like the ITU (International Telecommunication Union) [26], widely accepted AI standards would enable companies to scale innovations globally without contending with disparate national regulations.

Historical evidence suggests that flexibility, industry engagement, and iterative policy development are critical for regulating disruptive technologies while fostering growth. Rather than relying solely on centralized, rigid control, a combination of soft governance such as industry guidelines and targeted statutory intervention at critical junctures has yielded the most favorable outcomes. The challenge for AI governance is to implement measured, evidence based interventions informed by expert and industry input, while remaining adaptable as the technology and its impacts evolve. Such an approach can help ensure that AI contributes effectively to economic growth and societal welfare, consistent with lessons learned from past technological revolutions.

## VI. PRINCIPLE-BASED AI REGULATORY FRAMEWORK FOR ETHICAL AND INNOVATIVE GROWTH

As artificial intelligence becomes increasingly central to modern economies, a balanced regulatory framework is essential to ensure ethical, transparent, and effective governance while promoting innovation and capitalist growth. The following recommendations outline a principle-based approach that upholds safety and fairness without stifling technological advancement or economic competitiveness. This framework spans domestic oversight, international coordination, industry compliance, and risk mitigation. This framework is structured into five key steps with actionable guidance, ensuring AI governance is balanced, adaptable, and internationally aligned while fostering innovation.

### A. Balancing Oversight with Technological Advancement

*1) Focus on Outcome based Regulation:* A successful AI framework should be grounded in clear, high level principles that safeguard individuals and society while enabling innovation. Global guidelines such as those from the OECD and the EU emphasize outcomes including safety, fairness, transparency, accountability, privacy, and human oversight [27]. By focusing on these outcomes rather than rigid, technology specific rules, regulations can remain both neutral and flexible. In practice, this requires that AI developers ensure their systems are non-discriminatory, transparent regarding decision processes, secure against misuse, and subject to human review, while retaining flexibility in achieving these objectives.

*2) Implement Risk based Oversight:* Given the rapid pace of AI development, the regulatory framework must be adaptive. Mechanisms such as periodic reviews or sunset clauses can update guidelines as new capabilities and use cases emerge. For instance, lists of "high-risk" AI applications should be revisable in response to technological and societal changes [27]. An adaptive oversight body or dedicated AI office can issue rolling guidance and best practices as novel applications (e.g., generative AI or autonomous vehicles) arise, ensuring that governance evolves with innovation.

*3) Support Proportional Compliance for Startups and SMEs:* Clarity and proportionality are crucial to prevent burdensome compliance costs, particularly for startups and small-to-medium enterprises (SMEs). A risk-based approach should impose stricter requirements only on AI applications that pose higher potential harm, thereby avoiding overregulation of low-risk innovations [28]. To support SMEs, regulators should provide simplified compliance guidelines, templates, and accessible support resources (e.g., helplines or online FAQs). Engaging startups and SMEs in the rule-making process helps ensure that requirements remain practical and do not inadvertently stifle innovation.

*4) Introduce Regulatory Sandboxes for Innovation:* Regulatory sandboxes provide controlled environments where companies can test AI systems under relaxed regulatory conditions with close supervision. These sandboxes enable innovators to experiment with new solutions without the immediate burden of full compliance, while regulators gain early insights into emerging technologies [28]. Participation in such sandboxes, particularly with incentives for startups, allows for iterative learning and gradual scaling toward full regulatory compliance. This light-touch oversight model has been adopted in frameworks like the EU AI Act, proving effective in preventing harm without stifling innovation.

### B. Mechanisms for International Coordination

*1) Establish a Global AI Governance Body:* AI is a global technology, and its effective governance requires international coordination. A globally recognized AI governance body or consortium potentially under the auspices of the UN, G20, or an expanded mandate for the OECD should convene governments, industry experts, and civil society to develop common principles, share best practices, and resolve cross-border issues [27]. Functions of this body might include maintaining an international registry of high-risk AI systems, issuing ethical and safety guidelines, and mediating disputes. Aligning countries on core values (e.g., human rights, transparency) and shared definitions (e.g., what constitutes "high-risk" AI) reduces regulatory friction and creates a predictable environment for global operations.

*2) Encourage Harmonized Cross-Border Standards:* Adoption of cross-border standards for data governance, safety testing, and liability is essential. International standards organizations (e.g., ISO/IEC) and multi-country agreements can define baseline requirements that ensure compliance in one jurisdiction translates to others. For instance, establishing mutually recognized protocols for safety testing (including shared benchmark datasets and stress-test procedures) and consistent liability frameworks will protect companies globally and prevent regulatory arbitrage [27].

*3) Develop International Regulatory Sandboxes:* Building on domestic sandbox concepts, cross-border AI sandbox trials should be encouraged. Collaborative sandbox programs among multiple countries can allow AI solutions to be piloted under a unified oversight framework, which is particularly useful for applications operating internationally (e.g., global financial AI tools). These international sandboxes, overseen jointly by regulators from participating countries, facilitate simultaneous testing under agreed safeguards (e.g., privacy protection and continuous monitoring) and help inform the refinement of international standards [29].

### C. Industry Compliance and Responsible AI Innovation

*1) Encourage Self Regulatory Mechanisms with Accountability:* While governments set the rules, industry is key to implementing AI ethics and safety in practice. Companies should be encouraged to establish internal AI ethics committees or review boards, adopt industry codes of conduct, and integrate ethical risk assessments throughout the development lifecycle. To ensure accountability, these self-regulatory measures can be verifiable via independent third-party audits [30]. Regular audits can examine data handling, algorithmic fairness, security measures, and transparency practices. Establishing audit frameworks and certification schemes similar to ISO quality audits helps maintain robust checks and balances

*2) Provide Economic Incentives for Ethical AI:* Aligning market incentives with responsible AI development can be achieved by rewarding companies that prioritize ethics, transparency, and safety. Governments might introduce tax credits, grants, or expedited regulatory approvals for businesses that meet high ethical standards. Such incentives lower the cost of compliance and provide a competitive advantage to firms that market themselves as trustworthy AI providers. Public procurement policies that favor vendors with strong AI governance standards can further drive ethical practices.

*3) Standardize AI Impact Assessments (AIA):* Mandating or encouraging standardized AI impact assessments can help organizations evaluate the societal, ethical, and safety implications of their systems prior to deployment. An AIA framework should guide companies to assess factors such as the system's purpose, data usage, potential biases, privacy implications, and risk mitigation strategies. Periodic updates to these assessments—especially when systems are modified or expanded—will ensure continuous risk management. Standardized AIA protocols also provide valuable data for third-party audits and help establish industry benchmarks for best practices [28].

*D. AI Risk Assessment and Mitigation*

Not all AI systems carry the same level of risk; hence, a tiered, risk-based regulatory model is essential [31]. Under this model, AI applications are categorized into risk tiers, each with proportionate regulatory requirements:

In the Table II, each tier builds on the previous (higher tiers must also meet lower-tier requirements). This tiered approach follows the widely accepted principle that regulation should be proportionate to risk, focusing efforts where they matter most [31]. Low-risk innovations can proceed with minimal intervention, supporting rapid growth, whereas high risk applications face robust scrutiny to prevent harm. For high-risk AI, explainability and auditability are critical. These systems must be capable of providing understandable reasons for their outputs and maintain detailed documentation for independent review. Affected individuals should have the right to an explanation and a path to human recourse, thereby protecting against "black box" harm and ensuring alignment with legal and ethical norms [27].

*E. AI Incident Reporting & Global Adaptability*

*1) Establish an AI Incident Reporting System:* In addition to preventive measures, establishing an AI incident reporting system is crucial for continuous improvement. Such a system would require entities deploying AI—especially high-risk applications to report significant incidents or near-misses to a designated authority. Modeled after systems in aviation and cybersecurity, an AI Incident Reporting Hub could collect and analyze data to identify patterns of failure, inform regulatory updates, and guide industry best practices. Public transparency of incident data, where feasible, builds trust and facilitates preemptive actions to address emerging risks [32].

| Risk Tier | Description & Examples | Requirements |
|---|---|---|
| Tier 1: Low-Risk AI | AI applications with minimal or no potential for harm (e.g., scheduling tools, grammar correction, inventory optimization) | Light oversight, including basic adherence to data privacy and voluntary transparency practices |
| Tier 2: Moderate-Risk AI | AI systems that affect individuals but do not have life-altering consequences (e.g., customer service chatbots, recommendation engines) | Moderate oversight requiring transparency, periodic data and bias audits, and human fallback options for significant decisions. |
| Tier 3: High-Risk AI | AI with significant potential impacts on human rights, safety, or critical decision-making (e.g., healthcare diagnostics, autonomous driving, AI in hiring or credit decisions) | Strict oversight including pre-market assessments, mandated risk management systems, detailed documentation, third-party audits, and robust explainability. Human oversight is required in critical decision processes |
| Tier 4: Unacceptable or Prohibited AI | AI applications that pose extreme risks or violate fundamental rights (e.g., social scoring systems, autonomous weapons) | Such systems should be banned or heavily restricted, with enforcement through market bans and significant penalties |

TABLE II: Illustrative AI risk tiers with examples and corresponding regulatory measures

This principle based framework is designed to be actionable, scalable, and adaptable across jurisdictions. By following this five-step framework, AI governance remains proactive, ethical, and innovation-friendly. It ensures that 1) AI regulation evolves with technology 2) AI remains globally harmonized 3) Companies have clear compliance pathways 4) Riskier AI receives stricter oversight 5) AI innovation is encouraged while protecting society. Adopting these recommendations will help societies reap AI's benefits in a responsible, sustainable, and globally harmonized manner

## VII. CONCLUSION AND FUTURE WORK

In conclusion, effective AI governance will hinge on a collaborative, multi faceted approach that balances oversight with innovation. This paper's analysis highlights that global harmonization of AI principles and regulations is essential to avoid fragmented governance. International cooperation through bodies like the OECD, United Nations, and cross border agreements can establish common standards and interoperable frameworks that help every nation address AI's challenges cohesively. Such globally aligned principles not only prevent regulatory gaps but also foster trust and innovation, creating a stable environment in which AI can safely flourish. The future of AI governance holds significant opportunities if these strategies are pursued in unison. We may see the emergence of globally recognized AI governance bodies or treaties that set baseline rules for all (much like international agreements on climate or trade), providing clarity and confidence in AI systems across borders.

In summary, the governance of AI is entering a new era characterized by global unity of purpose, empowered self regulation, and regulatory creativity. By pursuing harmonized policies and standards globally, encouraging industry to lead in ethical AI practices, and adopting compliance models that evolve with technology, stakeholders can collectively ensure that AI's growth remains safe, equitable, and beneficial. This forward looking, collaborative governance approach will not only address the challenges of today but also future-proof the development of AI safeguarding society while unlocking AI's full potential for innovation and progress.

## VIII. Acknowledgment

This work is partially supported by the US National Science Foundation grant 2431531.